\documentclass[10pt,twocolumn,letterpaper]{article}

\usepackage{iccv}
\usepackage{times}
\usepackage{epsfig}
\usepackage{graphicx}
\usepackage{amsmath}
\usepackage{amssymb}
\usepackage{subfig}
\usepackage{slashbox}

\usepackage{hyperref}

\newcommand*\captiontype[1]{\def\@captype{#1}}

\iccvfinalcopy 

\ificcvfinal\fi
\begin{document}

\title{NODE: Extreme Low Light Raw Image Denoising using a Noise Decomposition Network}


\author{Hao Guan$\dagger$\\
{\tt\small gh279760559@gmail.com}\\
\and
Liu Liu\\
{\tt\small zoe.liuliu@huawei.com}
\and
Sean Moran\\
{\tt\small sean.j.moran@gmail.com}
\and
Fenglong Song\\
{\tt\small songfenglong@huawei.com}
\and
Gregory Slabaugh\\
{\tt\small greg.slabaugh@gmail.com}\\
\quad
\\
\hspace{-2.2in}Huawei Noah's Ark Lab\\
}


\twocolumn[{%
\renewcommand\twocolumn[1][]{#1}%
\maketitle
\centering
\includegraphics[width=0.99\textwidth]{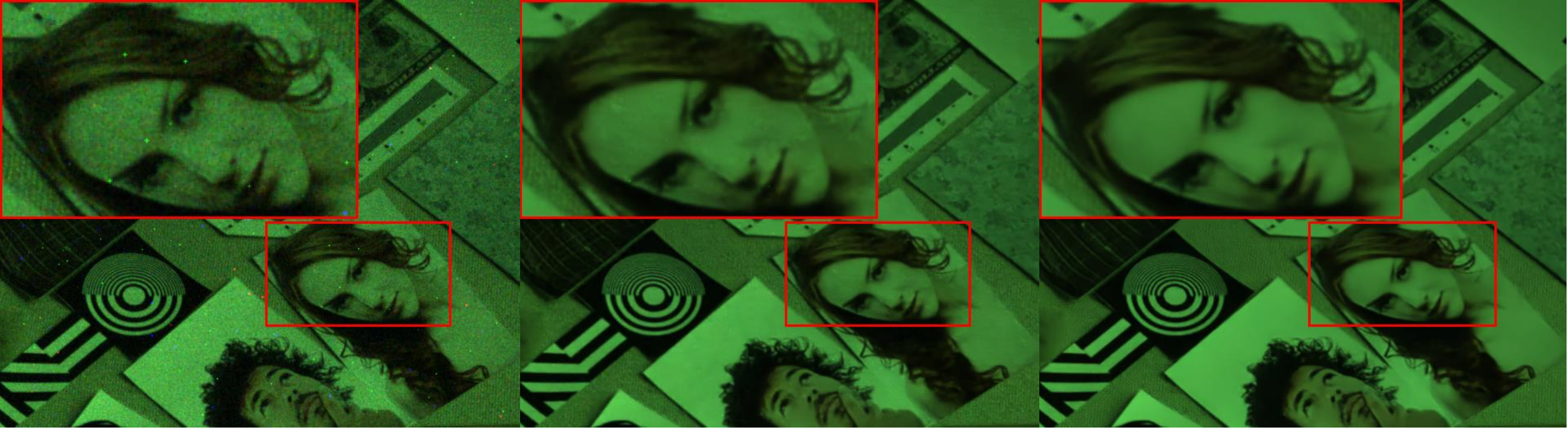}
\captionof{figure}{An example of extreme low light denoising. From left to right: the input image, the image denoised using DnCNN~\cite{zhang2017beyond} and denoised result using the proposed technique (NODE).  Note: for visualisation, the images have been demosaiced and brightened (but not white balanced).}
\vspace{3em}
\label{fig:example}
}
]

\begin{abstract}
\vspace*{-2em}
Denoising extreme low light images is a challenging task due to the high noise level. When the illumination is low, digital cameras increase the ISO (electronic gain) to amplify the brightness of captured data.  However, this in turn amplifies the noise, arising from read, shot, and defective pixel sources.  In the raw domain, read and shot noise are effectively modelled using Gaussian and Poisson distributions respectively, whereas defective pixels can be modeled with impulsive noise.  In extreme low light imaging, noise removal becomes a critical challenge to produce a high quality, detailed image with low noise. In this paper, we propose a multi-task deep neural network called \emph{\textbf{No}ise \textbf{De}composition (\textbf{NODE})} that explicitly and separately estimates defective pixel noise, in conjunction with Gaussian and Poisson noise, to denoise an extreme low light image.  Our network is purposely designed to work with raw data, for which the noise is more easily modeled before going through non-linear transformations in the image signal processing (ISP) pipeline.  Quantitative and qualitative evaluation show the proposed method to be more effective at denoising real raw images than state-of-the-art techniques. \let\thefootnote\relax\footnotetext{{$\dagger$} \text{Corresponding author}}

\end{abstract}

\section{Introduction}
Image denoising is a fundamental problem in computer vision and image processing. The task is to recover the latent clean image $\mathbf{x}$ given a noisy observation $\mathbf{y}$ in the presence of unknown additive noise $\mathbf{v}$, namely:
\begin{equation}
    \mathbf{y} = \mathbf{x} + \mathbf{v}
    \label{eq:additiveNoise}
\end{equation}
If the noise $\mathbf{v}$ can be effectively estimated, it can subtracted from noisy image $\mathbf{y}$ to produce a denoised result. Therefore, an effective approach to denoising is estimation of accurate, per-pixel noise.  However, in modern digital cameras this is a difficult challenge for the following reasons:

  \textbf{Non-linear ISP operations}: Current Image Signal Processing pipelines perform a large number of non-linear operations, including demosaicing and high dynamic range compression.  Consequently, noise that can be accurately characterised in the raw domain becomes very complicated at the end of the pipeline.  For tractability, much previous work on denoising makes the \emph{over}-simplifying assumption that the noise is white and Gaussian, and applies a denoiser to the RGB image produced by the ISP.  However, recent work~\cite{brooks18raw} shows the power of building denoising methods in the raw domain, where the modeled noise better
  matches the physics of image formation. 
  
  \textbf{Types of noise}: There are several types of noise present in raw data. \emph{Read noise} results from thermal and  electrical  noise in the electronics of the imaging sensor.  This stochastic noise can be modelled with a Gaussian distribution. \emph{Shot noise} is related to the number of photons arriving at the sensor, and is therefore brightness dependent.  Shot noise can be effectively modeled with a Poisson distribution. \emph{Defective pixel noise} arises from different sensitivities of pixels on the sensor to incoming light. This can occur as a consequence of the manufacturing process of the sensor, or due to random failures during service, so these sensitivities can change with time and operating condition.   Some pixels may become saturated (maximum brightness) or capture no light (minimum brightness) as well as levels inbetween.  A spatially impulsive noise model can represent this type of noise.
  
  \textbf{Light level (ISO)}:
  In digital cameras, ISO is used to adjust the light sensitivity of imaging sensors.  As scene lighting decreases, the ISO can be increased to apply an electronic gain to produce a correct exposure.  However, amplifying the signal also increases the noise.  Particularly in extreme low light scenes (ISO$>$3200), all noise sources described in the previous paragraph are increased, especially defective pixels, which become more apparent in extreme low light.  This has a detrimental effect on image quality.  


\begin{figure*}[t!]
\begin{center}
\includegraphics[width=0.8\textwidth]{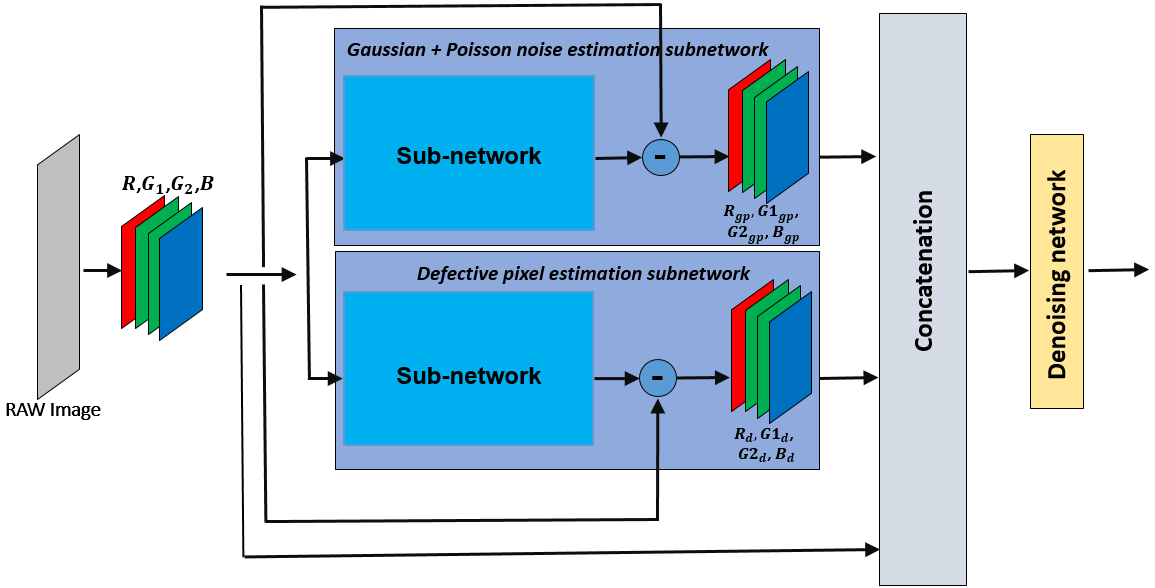}
\end{center}
   \caption{The full architecture of NODE, the proposed method.}
\label{fig:overall}
\end{figure*}

This paper addresses these challenges by proposing a novel image denoising network designed for extreme low light imaging.  The method, called \emph{\textbf{No}ise \textbf{De}composition (NODE)} works in the raw domain where the noise can be more accurately modeled~\cite{brooks18raw}. NODE has a multi-task design that decomposes the noise using separate Gaussian+Poisson and defective pixel noise estimators.  We demonstrate that NODE is more effective in extreme low light imaging compared to single-task state-of-the-art denoisers.  An example is shown in Figure~\ref{fig:example} comparing the denoised result produced using DnCNN~\cite{zhang2017beyond} to that of NODE.

\section{Related Work}

We briefly review traditional approaches to denoising as well as recent deep learning approaches, along with literature focussing on low light denoising. For more detailed coverage of the image denoising literature, we refer the interested reader to~\cite{Bertalmío2018Book}.

\subsection{Traditional methods}
Noise is a ubiquitous phenomena in  images dating back to the first photograph taken in the mid-1820s by Nic\`{e}phore Ni\`{e}pce.  Early denoising work using digital image processing techniques includes lowpass filters implemented in the spatial domain (box filters, Gaussian filters) or transform (Fourier, DCT) domain~\cite{Castleman1996Book}.  While effective to remove high frequency noise, these methods also tend to blur the image. Subsequent transformation domain work such as wavelet shrinkage~\cite{donoho95} better preserves detail by exploiting sparsity in the wavelet domain. Edge-preserving denoising was also approached using anisotropic diffusion~\cite{perona1990} and bilateral filtering~\cite{Tomasi1998}.  More recent approaches exploit self-similarity often inherent in an image. Such work includes non-local means~\cite{Buades2005} and BM3D~\cite{dabov2007image}, the latter being a block-matching technique that exploits aggregates similar patches and filters collaboratively in the wavelet domain.

\subsection{Deep learning approaches}

Closely related to this paper, deep neural networks have been proposed to address the image denoising task.  Jain and Seung~\cite{jain2009natural} design a multi-layer convolutional neural network for image denoising. Burger \etal\cite{burger2012image} solve the denoising problem by training a multi-layer perceptrion (MLP). This paper shows than an MLP can achieve  comparable performance with BM3D~\cite{dabov2007image}. Xie \etal~\cite{xie2012image} combine sparse coding and a deep neural network pre-trained with denosing auto-encoder to solve denoising and inpainting problems. The architecture proposed by Zhang \etal~\cite{zhang2017beyond} learns the residual noise given a noisy image by combining convolution, batch normalization~\cite{ioffe2015batch} and rectified linear unit (ReLU)~\cite{nair2010rectified}. Their network, DnCNN, achieves better quantitative results, i.e. peak signal-to-noise ratio (PSNR) than the conventional state-of-the-art approaches~\cite{gu2014weighted, zoran2011learning, dabov2007image}. 

Ronneberger \etal~\cite{ronneberger2015u} propose an encoding-decoding framework for segmentation of biomedical images. This paper utilizes skip connections, which have been applied to train very deep networks in many other applications~\cite{isola2017image, cciccek20163d, jin2017deep, wang2018high, chen2017photographic}. Mao \etal~\cite{mao2016image} propose an encoder-decoder style network (RED) for image restoration. This method demonstrates that using nested skip connections, the training process converges more easily and quickly.

Motivated by the persistence from human thought, Tai \etal~\cite{tai2017memnet} introduced deep persistent memory neural network called MemNet for image restoration. In this network, the memory block  contains a recursive unit and a gate unit to learn multi-level representations with different receptive fields through an adaptive learning process. The paper shows that combining short-term memory from the recursive unit and the long-term memory from the memory block, the method can resolve long term dependencies.
 

\subsection{Low light denoising}
Low light imaging is a challenging task that has received some attention in the literature.  Guo \etal~\cite{guo2017} present a Retinex-based method to adjust image brightness and allow for denoising of low light images. Chatterjee \etal present a joint denoising and demosaicking method~\cite{chatterjee2011noise} applied to low-light images, using vector upsampling based on Local Linear Embedding. While effective, residual noise is still present in the processed images. Li \etal~\cite{li2015low} explore a dark channel prior by inverting the image's brightness and applying BM3D-style denoising to superpixels. Remez \etal adaptively employ a deep neural network for Poisson noise removal on low light images~\cite{remez2017deep}.  

All of the methods described above are working in medium to low light. However, in this paper we address extreme-low light (ISO$>$3200) where denoising becomes more difficult. Chen \etal.~\cite{chen2018SID} propose a UNet-style network that learns the entire ISP pipeline for extreme low light environments.  In contrast, the proposed method focusses solely on the denoising task in the raw domain.

\subsection*{Our contributions}

While it is possible to build a single network to denoise an extreme low light image, the different noise sources confuse the denoiser, resulting in images that typically have undesirable residual noise. Critically, none of the prior work listed above specifically addresses defective pixel noise, which becomes more acute in extreme low light scenes.  

The key idea of this paper is to decompose the noise into Gaussian + Poisson noise and defective pixel noise, using dedicated sub-networks trained in a multi-task setting.  One task estimates the Gaussian + Poisson noise, and the other task estimates the defective pixel noise.  As these noise types are fundamentally different, each task can focus on a particular type of noise, producing a better result than training a single-task denoising network. 

Each sub-network is pre-trained with synthesized data. One sub-network is trained to estimate Gaussian+Poisson noise, whereas the other is trained to estimated defective pixel noise. These two noise estimates are then provided to a denoiser, as a concatenated input with the noisy image. The entire network is then fine-tuned on real images. In essence, NODE estimates the noise in an image-adaptive way, and then denoises the image.  The main contributions of the proposed methods are:

\textbf{Multi-task noise estimation}:  Our designed neural network can simultaneously estimate Gaussian+Poisson noise and defective pixel noise using two separate sub-networks.  This way, different parts of the network focus on specific types of noise.  To our knowledge, the defective pixel noise removal using a deep neural network has never done before.

\textbf{End-to-end training}:  The network is trained end-to-end using noisy and clean image pairs. NODE can produce an optimal solution to the problem of denoising of images corrupted with Gaussian + Poisson and defective pixel noise.

\textbf{Sub-network design}:  We design variants of UNet~\cite{ronneberger2015u} by replacing some maxpooling and deconvolution operations with shuffle operations (space to depth and depth to space).  While simple, these modifications help particularly with impulsive noise resulting from defective pixels.

\textbf{Extreme low light denoising}: The proposed method is designed to work in extreme low light scenarios and is shown to be more effective than conventional denoising techniques. 

\begin{figure*}[t!]
\begin{center}
\includegraphics[width=0.85\textwidth]{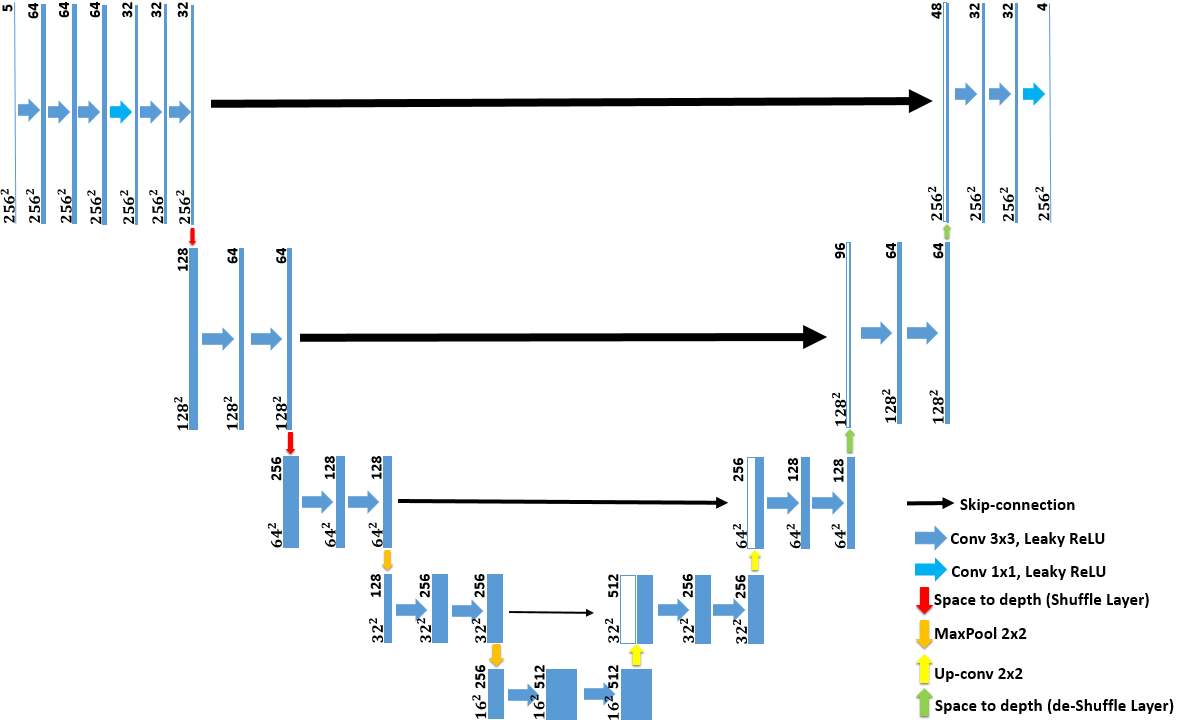}
\end{center}
   \caption{The architecture of sub-network.}
\label{fig:subnetwork}
\end{figure*}

\section{NODE: Noise Decomposition Network}

In this section, we present details of the decomposition network and explain how it works with the sub-network.

\subsection{The Overall Architecture}

The overall framework for NODE is presented in Figure~\ref{fig:overall}.  First, the method takes the raw noisy image from the Bayer pattern (consisting of $R$, $G_1$, $G_2$, and $B$ pixels) and packs it into four channels for subsequent processing. These channels are at half the width and height of the original image.  Packing is necessary to group same-color pixels together for subsequent convolutional layers.

The packed, noisy image is then input into two subnetworks: a Gaussian + Poisson noise estimation sub-network, and a defective pixel estimation sub-network. This multi-task architecture is designed to separately decompose the noise, allowing each subnetwork to focus on a different task as the noise types are very different.  Gaussian+Poisson noise corrupts every pixel in the image, whereas defective pixel noise is spatially sparse, affecting only certain pixels (which may vary over time). By decomposing the noise into these two different streams, NODE can achieve better results than a single-task denoising network as our experimental results will show.

The upper branch in the NODE architecture estimates a packed form of Gaussian+Poisson noise, essentially the predicted noise at each pixel, but packed into four channels (corresponding to $R$, $G_1$, $G_2$, and $B$).  Similarly, the lower branch in the NODE architecture estimates a packed form of the defective pixel noise.  These eight packed estimated noise channels serve as an initial estimate of the noise in the raw image.  While it would be possible to directly subtract these noise sources from the raw image to produced a denoised output, instead, NODE concatenates the estimated noise with the noisy raw image and feeds a 12 channel input to a denoising subnetwork.  This final network refines the noise estimate to produce the final denoised image. 

Pre-training is critical to the success of the Gaussian+Poisson and the defective pixel estimation subnetworks. Each subnetwork is pre-trained using synthetic data of a specific type.  For example, in the Gaussian+Poisson subnetwork, we start with clean raw images, and degrade them by adding Gaussian+Poisson noise.  We then train the subnetwork to learn how to predict a clean image given a noisy image degraded by Gaussian+Poisson noise.  A similar process is performed for the defective pixel subnetwork.

Let the noisy input image be denoted as $\mathbf{y}$, the per-pixel Gaussian+Poisson noise as $\mathbf{v}_{GP}$ the per-pixel defective pixel noise $\mathbf{v}_D$, and the clean image as $\mathbf{x}$.  In this case, we can rewrite Equation~\ref{eq:additiveNoise} as
\begin{equation}
    \mathbf{y} = \mathbf{x} + \mathbf{v}_{GP} + \mathbf{v}_{D}
\end{equation}
The upper subnetwork is trained to remove Gaussian+Poisson noise, so given a noisy image $\mathbf{y}$ it regresses an image $R(\mathbf{y}; \mathbf{\theta}) = \mathbf{\hat{x}} + \mathbf{\hat{v}}_{D}$, where $\mathbf{\hat{x}}$ and $\mathbf{\hat{v}}_{D}$ are estimated clean image plus defective pixel noise.  This result is subtracted from the original noisy input to produce the estimated per-pixel Gaussian+Poisson noise, i.e.
\begin{eqnarray}
    \mathbf{y} - R(\mathbf{y}; \mathbf{\theta}) & = & \left(\mathbf{x} + \mathbf{v}_{GP} + \mathbf{v}_{D}\right) - \left(\mathbf{\hat{x}} + \mathbf{\hat{v}}_{D}\right)  \\
     & \approx & \mathbf{\hat{v}}_{GP} 
\end{eqnarray}
This is the output of the upper subnetwork and is represented by the four channels $\mathrm{R_{gp}}, \mathrm{G1_{gp}}, \mathrm{G2_{gp}}, \mathrm{B_{gp}}$ in Figure~\ref{fig:overall}.  In a similar fashion, the lower defective subpixel estimation subnetwork estimates the per-pixel defective pixel noise, $\mathrm{R_{d}}, \mathrm{G1_{d}}, \mathrm{G2_{d}}, \mathrm{B_{d}}$.

With the pre-trained subnetworks in place, the entire NODE architecture is fine-tuned, end-to-end on real images.  This process adapts the weights learned from synthetic noise to that of real images. In this way, the method first adaptively estimates the noise at each pixel, and then applies a denoising network given the estimated noise. 

At inference, the input is a real noisy image, containing Gaussian + Poisson and defective pixel noise.  The noise is estimated, and concatenated with the original image for subsequent refinement by the denoising network, which produces the final denoised image.

Note that although it would be possible to directly subtract estimated Gaussian + Poisson and defective pixel noises from input images, NODE instead concatenates the estimated noise with the real image to refine the estimated noise.  This design can be thought of as an image-adaptive noise estimation, followed by a refinement denoising operation.




\subsection{The sub-networks}

Many neural network designs could be used for the sub-networks described above. In our experiments, we use an encoder/decoder network design inspired by UNet~\cite{ronneberger2015u}.  Our subnetwork architecture is in Figure~\ref{fig:subnetwork}. 

On the encoding path (starting from the upper left in Figure~\ref{fig:subnetwork}), a series of  convolutional layers with leaky ReLU~\cite{maas2013rectifier} extract features at high resolution, which is important in the raw domain as noise varies from pixel to pixel. We include a bottleneck layer for calculation efficiency. In practice, we find that these convolutional layers help preserve the high frequency detail. 

Next, the resolution of the image data is progressively reduced using a shuffle layer (red arrows in Fig.~\ref{fig:subnetwork}).  This layer reshapes the data so that the spatial resolution is decreased by a factor of two in width and height, but creates four times as many channels (space to depth).  Consequently, this shuffle layer makes the image size smaller while retaining important perceptual information. Symmetrically, deshuffle layers (green arrows) are used for the decoding process.  The shuffle and deshuffle layers are rendered with red and light green arrows in Figure~\ref{fig:subnetwork}. 

In subsequent processing, the resolution of the image data is progressively reduced using max pooling (golden arrow) is applied between each layer to allow more efficient subsequent processing.  On the decoding side, transposed convolution (up-conv, yellow arrow) is used for upsampling.  Skip connections are provided feed-through between corresponding layers as a effective way to configure the models to achieve good trainability and restore high frequency details.

\subsection{The Denoising Network}

A very similar architecture to the sub-network is used for the denoising network. The only differences are that
there are no extra convolutional layers at the beginning of the encoding path and there is only one pair of shuffling layer and deshuffling layer at the highest resolution replaced by the shuffle/unshuffling layers.
Please see the supplementary material more for a figure showing the denoiser network architecture.

\section{Implementation Details}
For this paper, we collected a new dataset using a Huawei P20 cellphone at ISO 12800.  The data is captured in a lossless, raw format using an RGGB Bayer pattern color filter array. At each pixel there is only a red, green, or blue color. 

\subsection{Synthetic Images}

\begin{figure*}[htb]
\centering
\begin{center}
\includegraphics[width=1\textwidth]{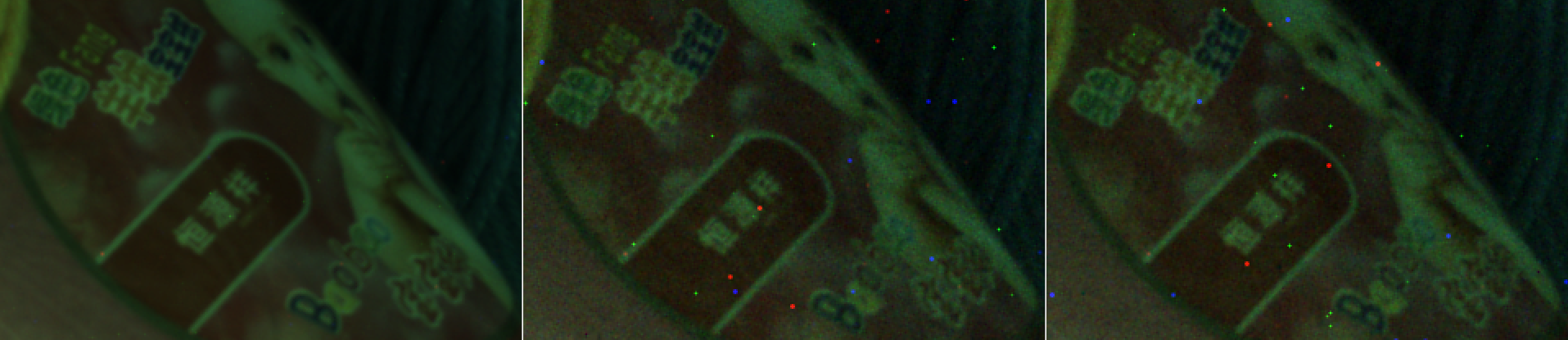}
\end{center}
\caption{The example of realistic noise synthesis using the Gaussian+Poission and defective pixel noise models. From left to right: a clean image, a real noisy image, and a synthesized image produced by adding noise to the clean image using the noise model.  Please zoom in for details.}
\label{fig:synthetic}
\end{figure*}

An important part of this work is noise synthesis to pre-train the sub-networks. For this, we first fit Gaussian+Poisson and defective pixel noise models to real data captured by the device, using a sequence of 12 images captured in a low light, static scene with a static camera.  We average the 12 frames, producing a mean image, which serves as a noise-free estimate $\mathbf{\bar{x}}$, and a variance image computed for each pixel $p$ across the sequence. The well known~\cite{Healey1994, Mildenhall2018} noise model $y_p \sim N(\bar{x}_p, \sigma_R^2 + \sigma_S \bar{x}_p)$, where $\sigma_R^2$ is the read noise, and $\sigma_S$ is the shot noise, respectively is fit to the noise variance as a function of intensity using RANSAC~\cite{fischler1981random} to robustly handle outliers in the data. Once fit, we can characterise noise using the Gaussian + Possion noise model.  Any pixels that exhibit noise inconsistent with this Gaussian + Poisson noise model is considered as defective pixel noise.  For this, we consider all pixels intensities outside of the 99\% confidence interval of the noise Gaussian + Poisson distribution as the defective pixels. Once the noise models are computed, we can then synthesize realistic noise for the device.  An example is shown in Figure~\ref{fig:synthetic}. In practice, we use 187 noisy sequences of 12 frames at high resolution (2736 $\times$ 3648) to generate the noise model and 145 images at the same resolution to generate two synthetic datasets which contain Gaussian + Poisson noise and Defective noise respectively.


\begin{table*}[htb]
\centering
\begin{tabular}{|c|cccccc|}
\hline
\backslashbox{Evaluation}{Methods} &BM3D~\cite{dabov2007image} &DnCNN~\cite{zhang2017beyond} &Unet~\cite{ronneberger2015u}  & MemNet~\cite{tai2017memnet} &RED~\cite{mao2016image} &Ours \\ 
\hline
PSNR $\uparrow$ (higher is better) &38.93 & 40.25  & 40.37  & 40.04  & 40.93  & \textbf{41.10}  \\
SSIM~\cite{wang2004image} $\uparrow$(higher is better)&0.9452 & 0.9770 & 0.9755 & 0.9763 & 0.9784 & \textbf{0.9789} \\
PSNR(MASK) $\uparrow$ (higher is better) &39.01 & 40.34  & 40.48  & 40.11  & 41.04 & \textbf{41.55}  \\
SSIM(MASK)~\cite{wang2004image} $\uparrow$(higher is better)&0.9487 & 0.9765 & 0.9752 & 0.9760 & 0.9780 & \textbf{0.9796} \\
PI~\cite{wang2018esrgan} $\downarrow$(lower is better)&6.5607 & 6.4801 & 6.5367 & 6.2536 & 6.4676 & \textbf{6.1065} \\ 
\hline
\end{tabular}
\caption{Quantitative performance comparing state-of-the-art single task denoisers retrained on the dataset to our proposed method.  Note: all testing images used to produce these results were heldout and independent of any training data.}
\label{tab}
\end{table*}

\section{Experiments}

We use the noise model to synthesize two different datasets, i.e. one dataset containing only the Gaussian + Possion noise model and the other containing only defective pixel noise. Then these two datasets are used for each subnet pre-training.  After pre-training, we then put the networks together into the full architecture of Figure~\ref{fig:overall} and fine-tune, training end-to-end on real data. 

\subsection{Experimental Setting}
For training the overall architecture, we collected 123 short/long exposure pairs at high resolution (2736 $\times$ 3648) using five Huawei P20 cellphones. The data was randomly split into training (90\%) and independent testing (10\%) sets with phones used in testing different from those in training. The training data was augmented by either flipping right/left or top/bottom or both. All the images are captured in an extreme-low light environment at ISO 12800. The network is trained using the Adam Optimizer~\cite{kingma2014adam} with $\beta_1=0.9$. We set the patch size $= 512 \times 512$, batch size $= 24$ and the learning rate  $= 1\mathrm{e}^{-4}$. We use $\mathrm{L_1}$ as the loss function.

The 10 MP images used in our study are real images taken from a phone, and our method can process them using a standard NVidia GTX 1080Ti GPU.  However, MemNet and RED were not able to process the full resolution images.  Therefore, we cropped the test images to a size $912 \times 1216$ for inference for fair comparison. Cropping was performed in the center of the image where there was the most salient content. We compare to state-of-the-art denoisers including BM3D~\cite{dabov2007image}, DnCNN ~\cite{zhang2017beyond}, MemNet~\cite{tai2017memnet}, RED~\cite{mao2016image} and Unet~\cite{ronneberger2015u}. Through a grid search, we set $\sigma=5$ for BM3D~\cite{dabov2007image} as it returned the best PSNR and SSIM~\cite{wang2004image} values. Aside from BM3D, all methods are implemented using Tensorflow~\cite{abadi2016tensorflow} and trained using the same dataset described above for the purpose of fair comparison. The settings of these competing methods are from the their papers respectively.

\subsection{Performance Evaluation}
\begin{figure*}[htb]
\centering
\subfloat[input]{{\includegraphics[width=0.47\textwidth]{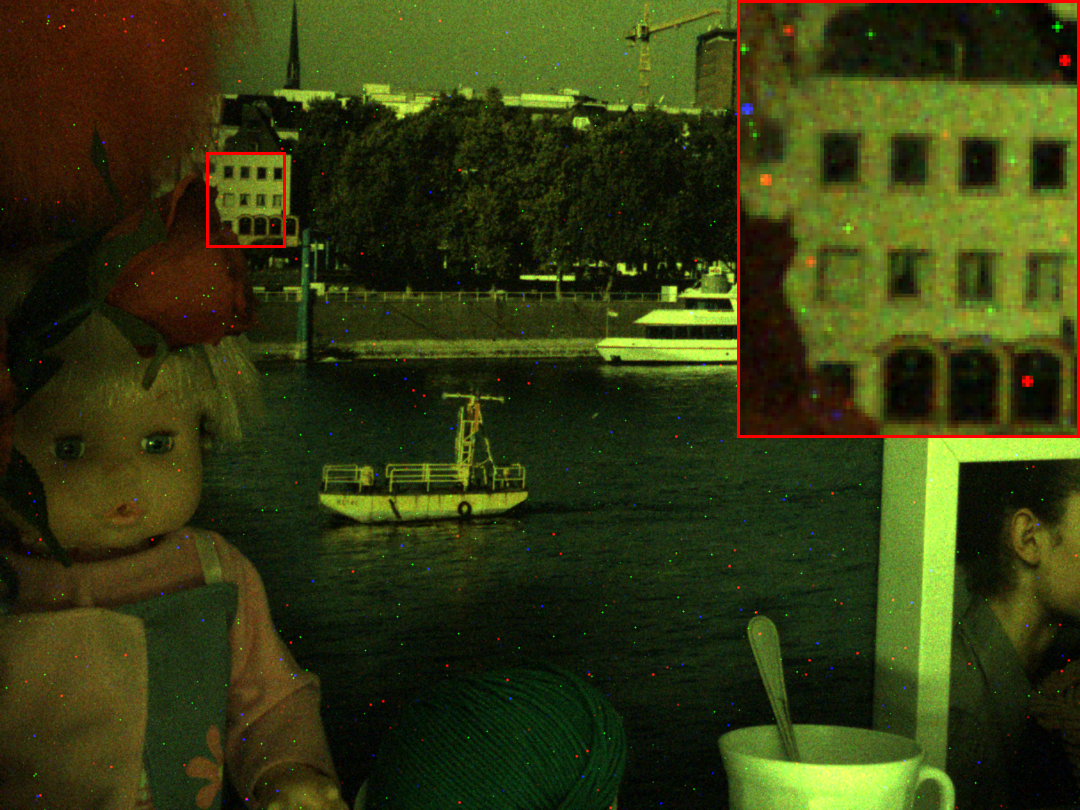}}}

\subfloat[DnCNN~\cite{zhang2017beyond}]{{\includegraphics[width=0.33\textwidth]{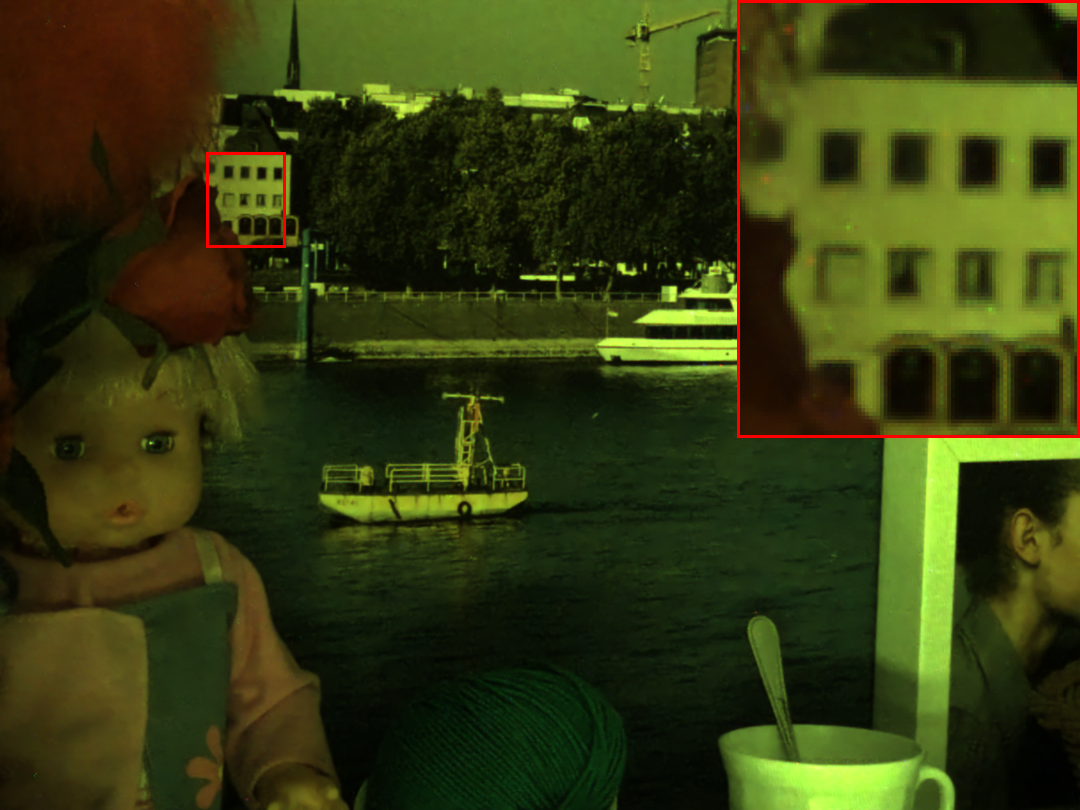}}}
\hfill
\subfloat[RED~\cite{mao2016image}]{{\includegraphics[width=0.33\textwidth]{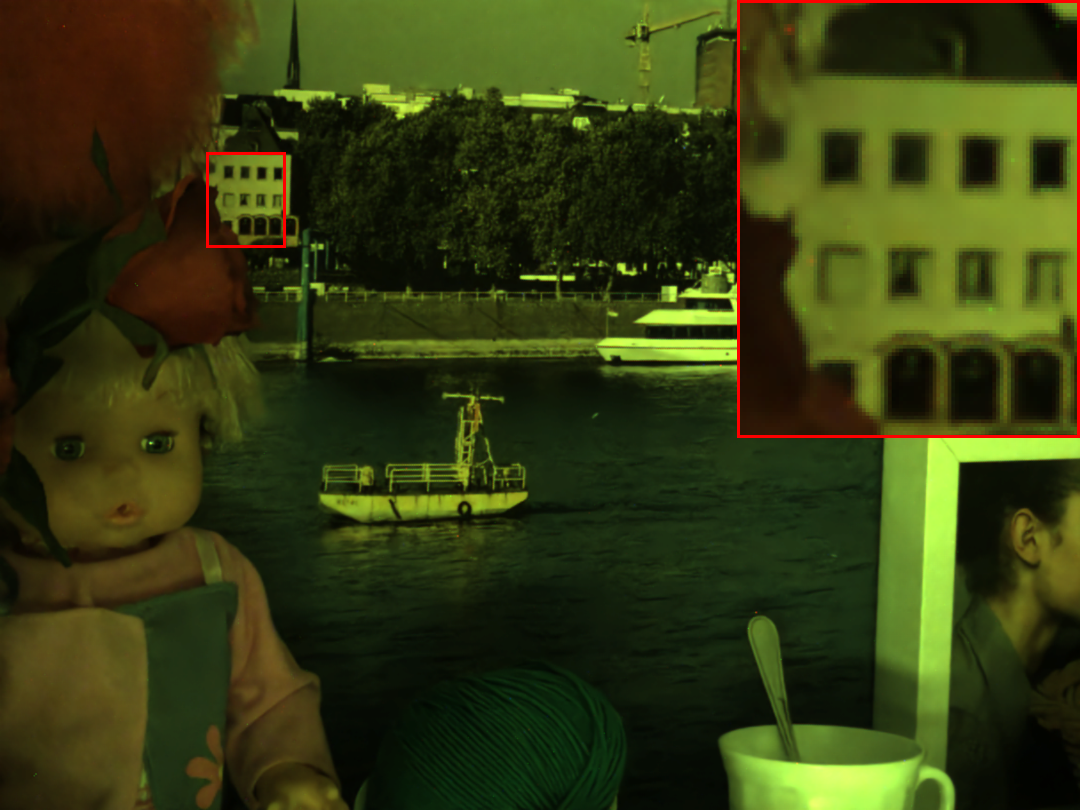}}}
\hfill
\subfloat[Unet~\cite{ronneberger2015u}]{{\includegraphics[width=0.33\textwidth]{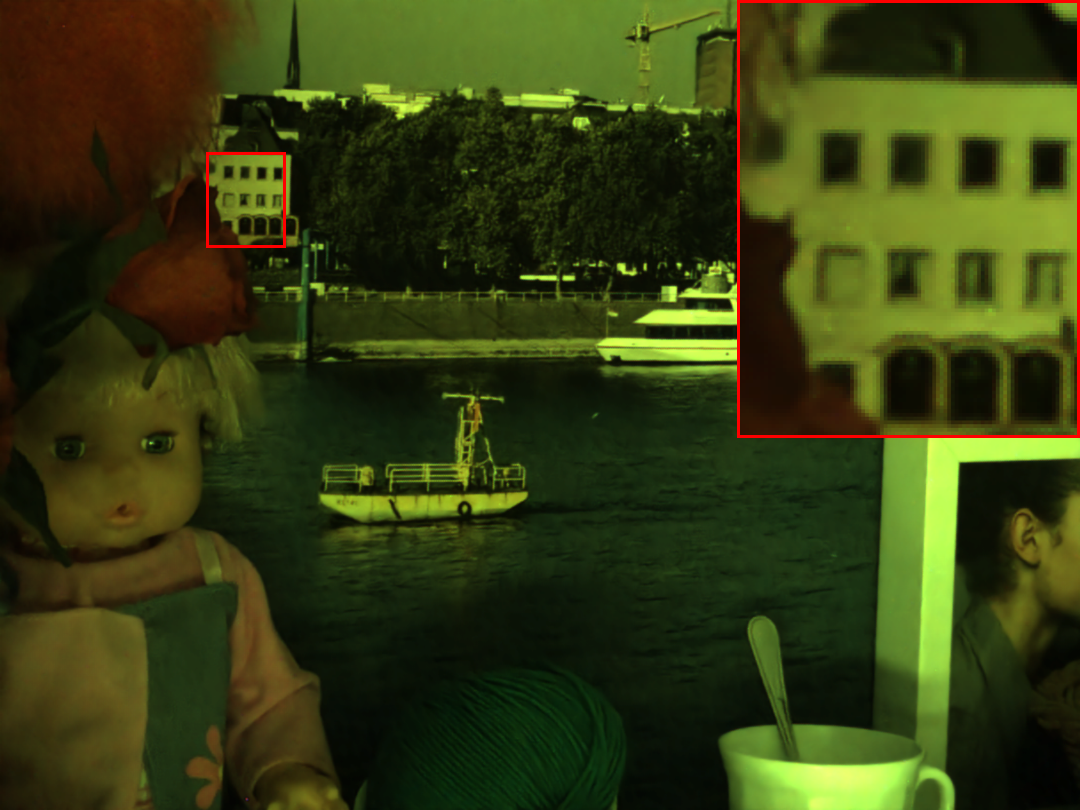}}}
\hfill
\subfloat[MemNet~\cite{tai2017memnet}]{{\includegraphics[width=0.33\textwidth]{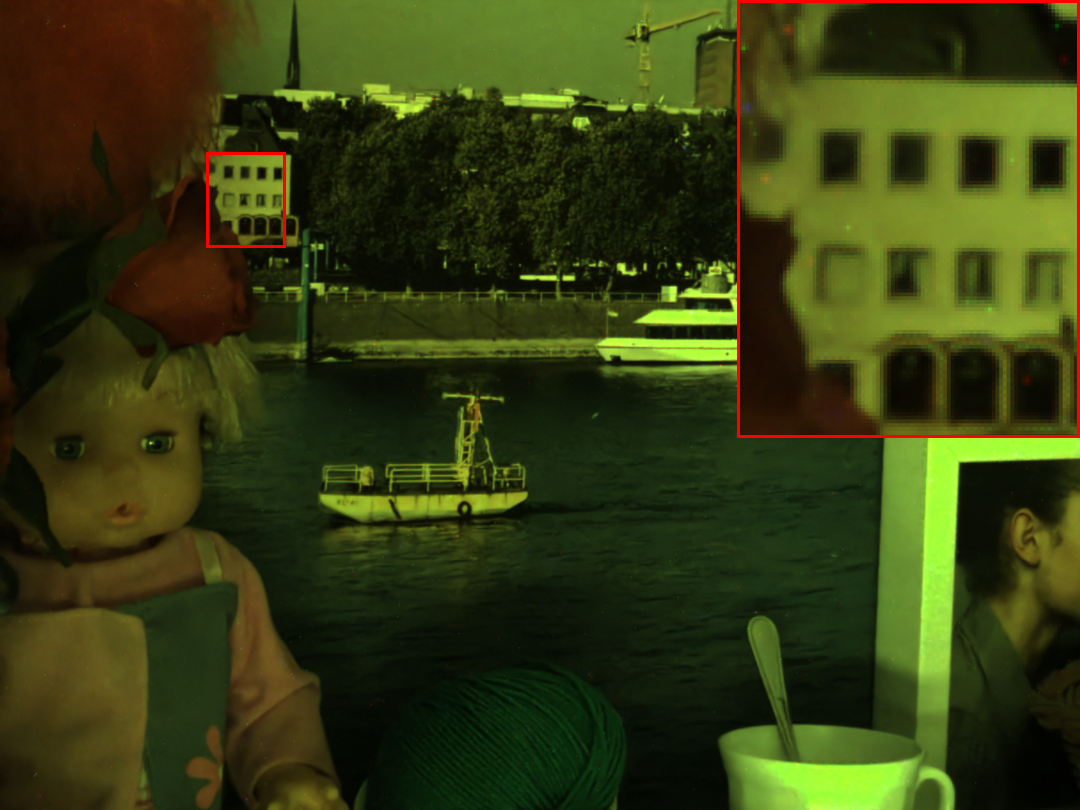}}}
\hfill
\subfloat[BM3D~\cite{dabov2007image}]{{\includegraphics[width=0.33\textwidth]{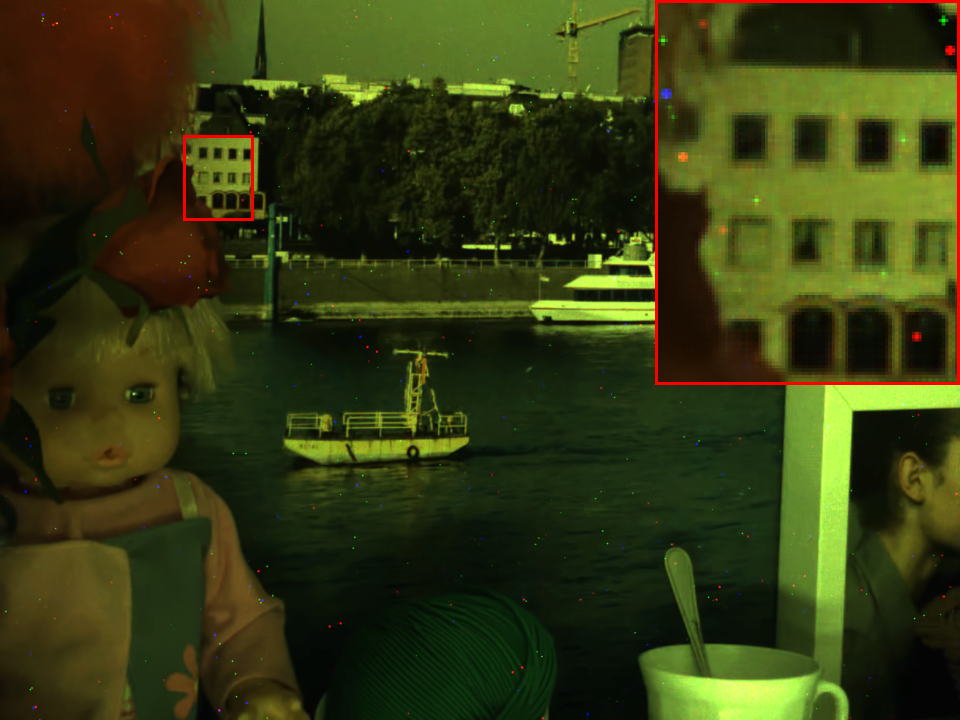}}}
\hfill
\subfloat[NODE]{{\includegraphics[width=0.33\textwidth]{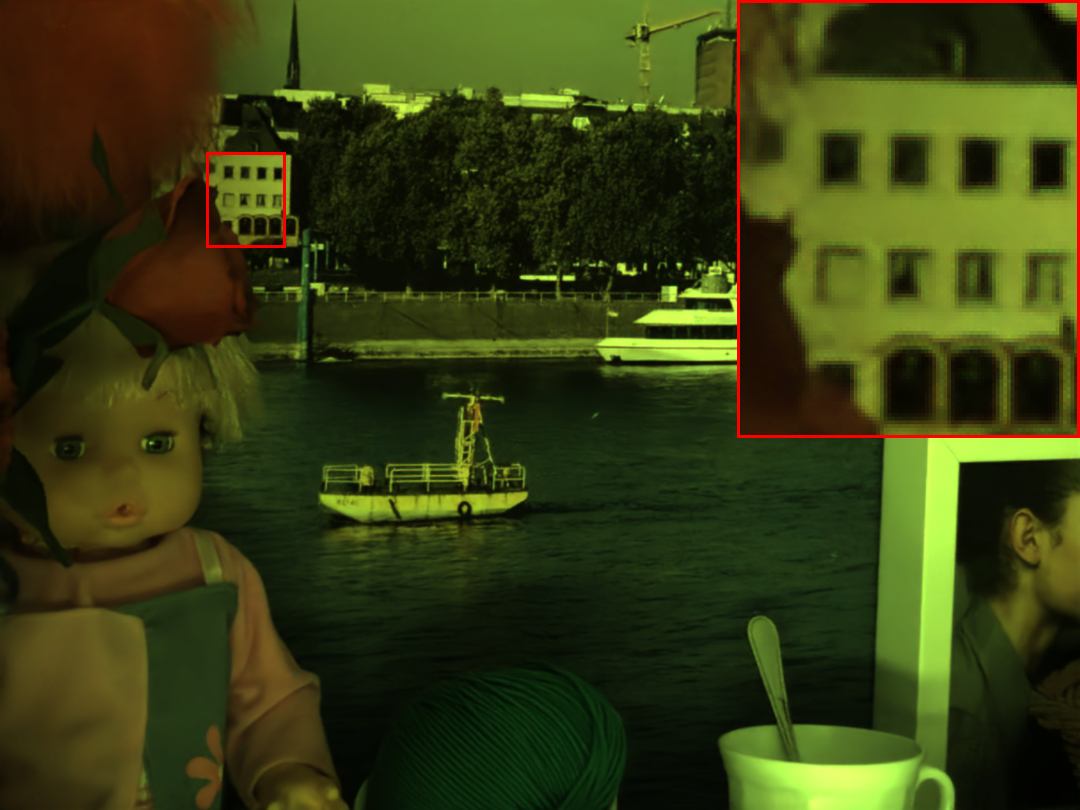}}}
\caption{Qualitative results comparing the proposed NODE method to the state-of-the-art.   Note that NODE is most effective at removing the influence of defective pixels, whilst at the same time removing the Gaussian+Poisson noise to deliver a high quality denoised image (please zoom in for details).}
\label{fig:qualitative1}
\end{figure*}

\begin{figure*}[htb]
\centering
\subfloat[input]{{\includegraphics[width=0.47\textwidth]{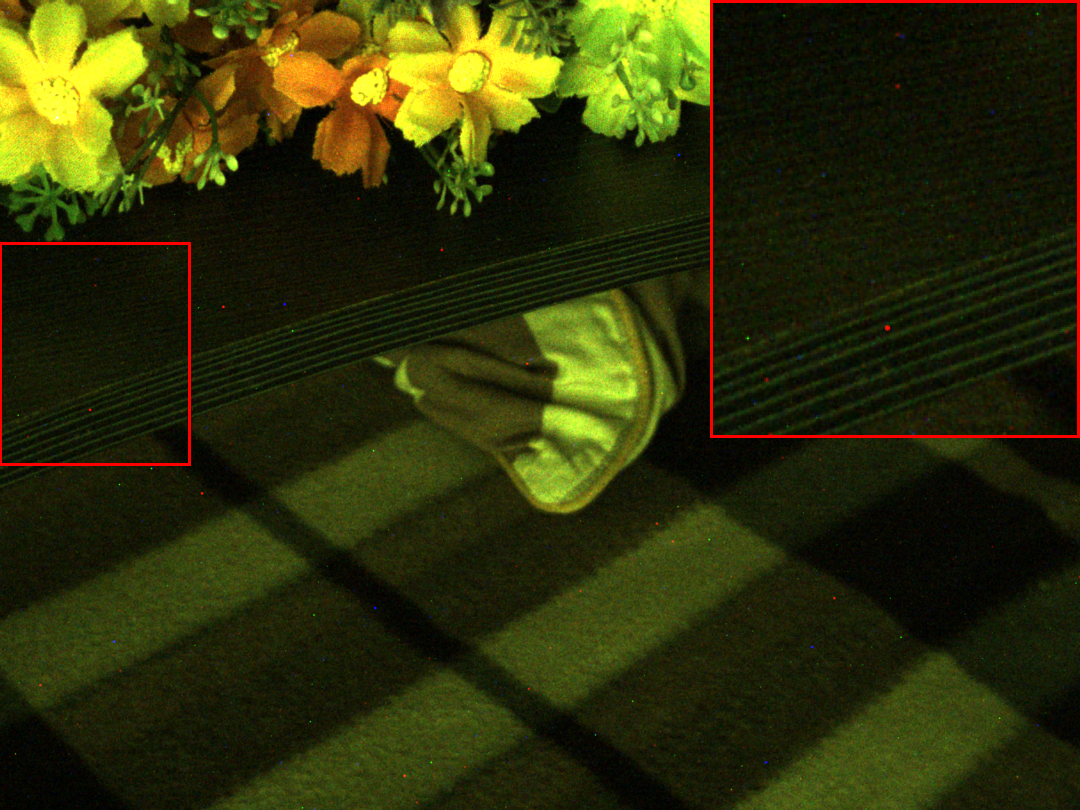}}}

\subfloat[DnCNN~\cite{zhang2017beyond}]{{\includegraphics[width=0.33\textwidth]{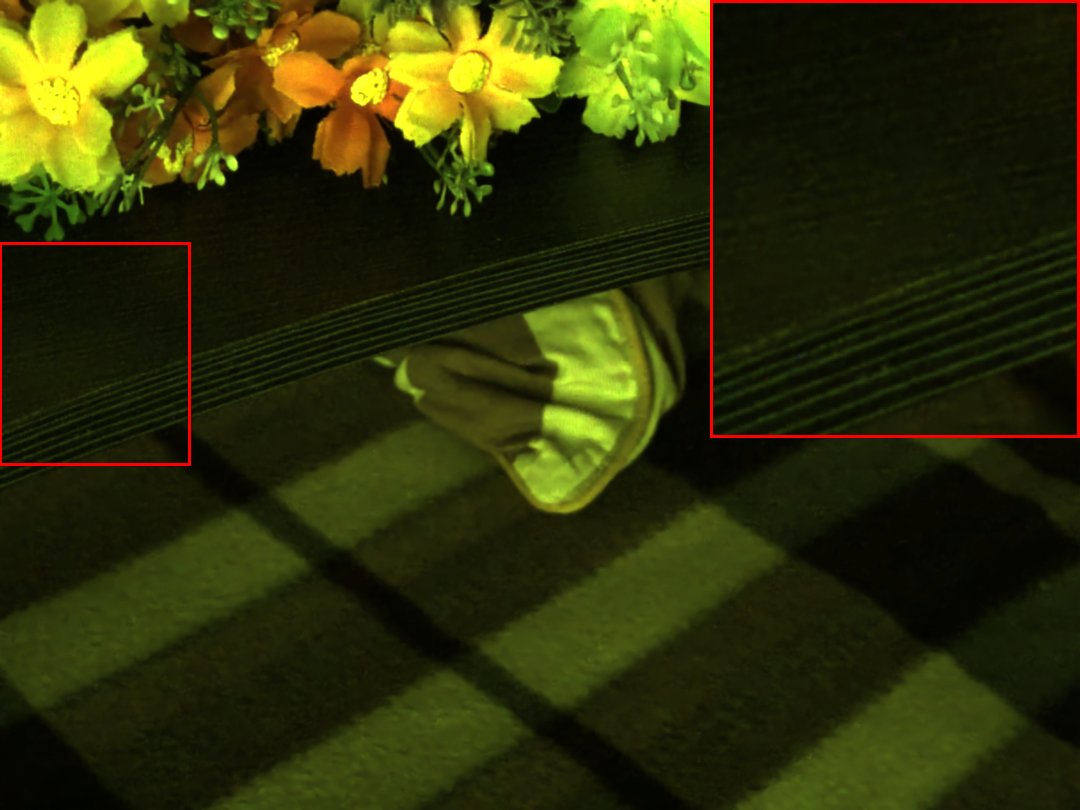}}}
\hfill
\subfloat[RED~\cite{mao2016image}]{{\includegraphics[width=0.33\textwidth]{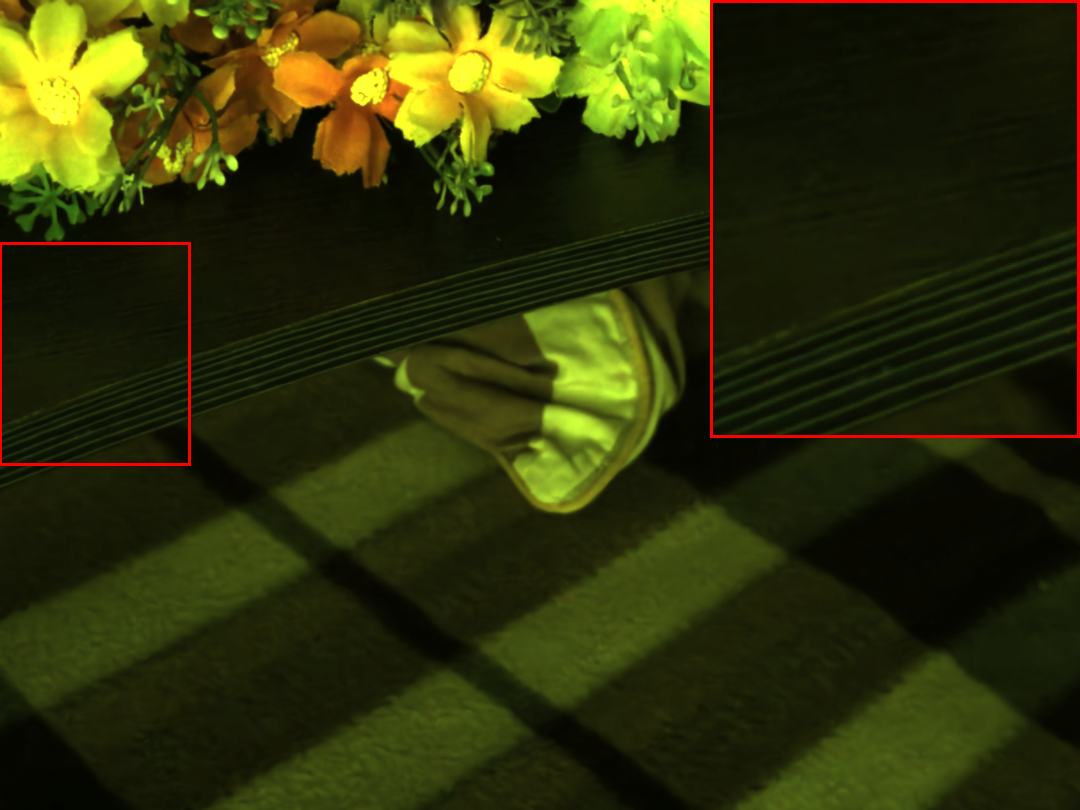}}}
\hfill
\subfloat[Unet~\cite{ronneberger2015u}]{{\includegraphics[width=0.33\textwidth]{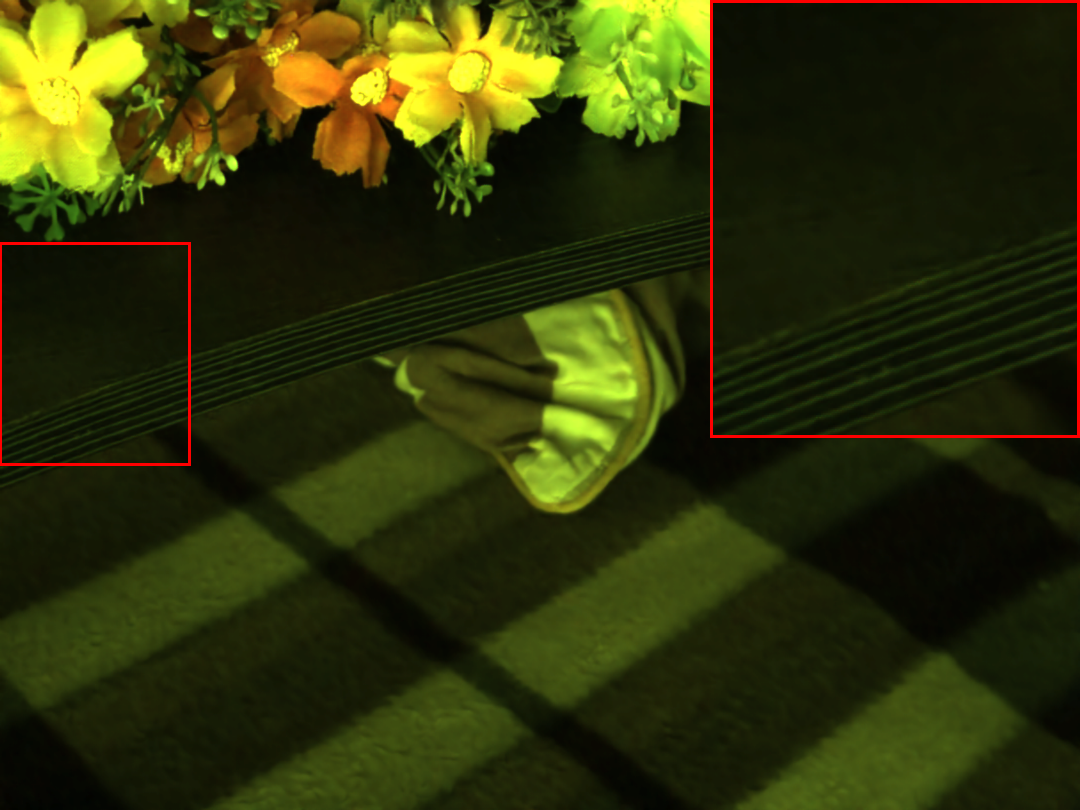}}}
\hfill
\subfloat[MemNet~\cite{tai2017memnet}]{{\includegraphics[width=0.33\textwidth]{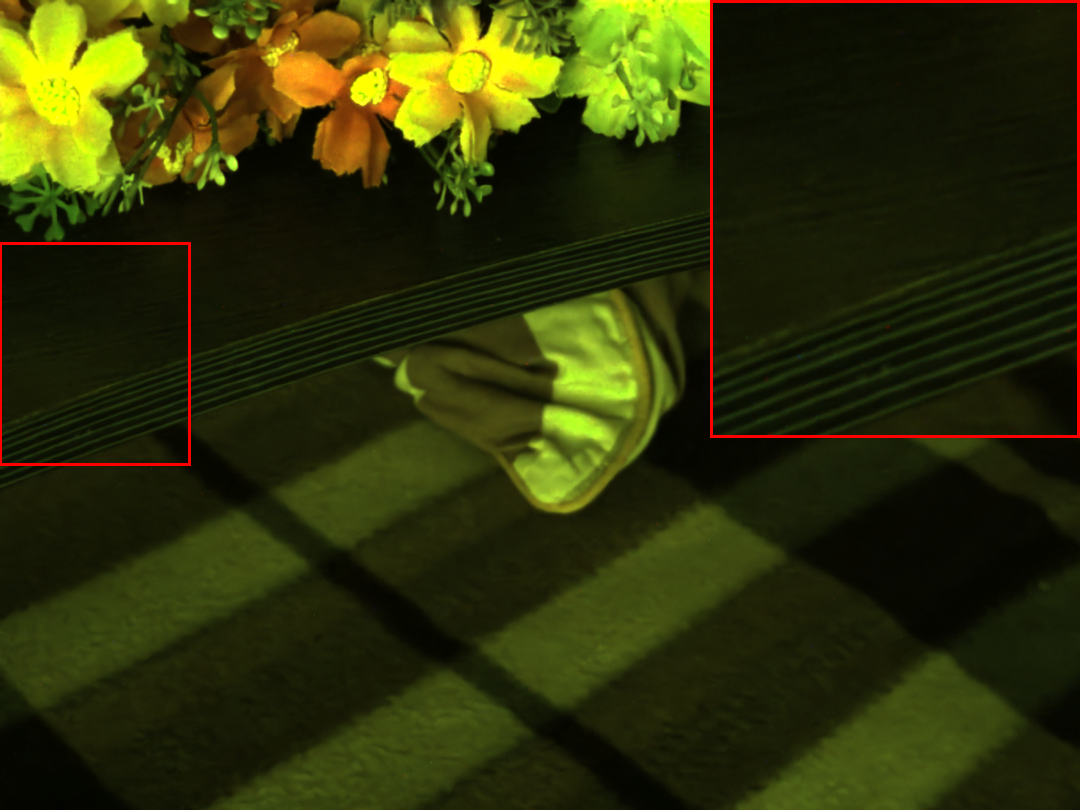}}}
\hfill
\subfloat[BM3D~\cite{dabov2007image}]{{\includegraphics[width=0.33\textwidth]{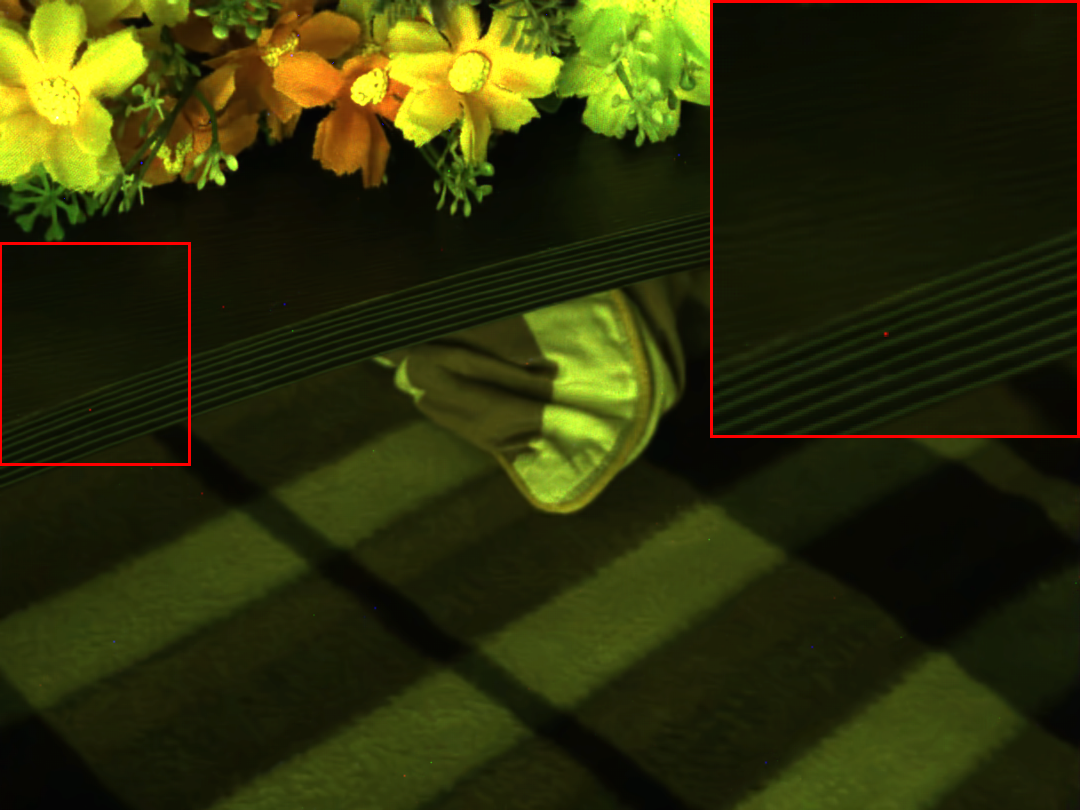}}}
\hfill
\subfloat[NODE]{{\includegraphics[width=0.33\textwidth]{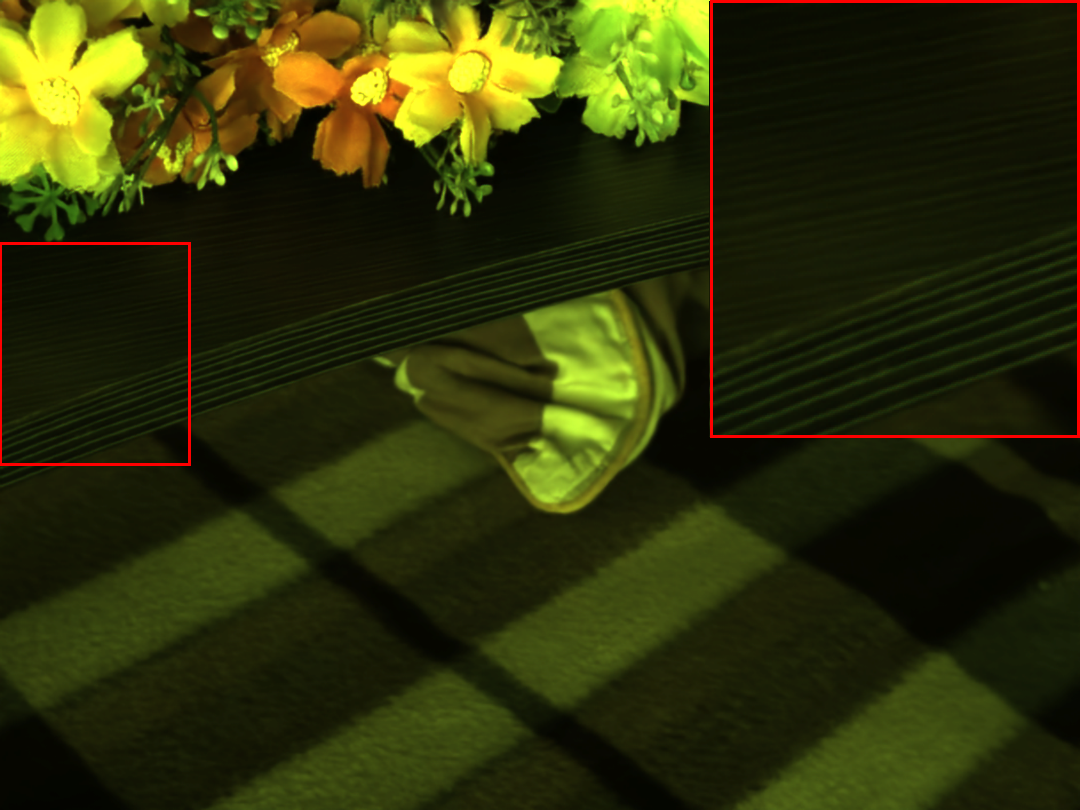}}}
\caption{Additional qualitative results comparing the proposed NODE method to the state-of-the-art.  Note that NODE is the only method that retains the fine texture on the table top surface  (please zoom in for details).}
\label{fig:qualitative2}
\end{figure*}

We evaluate the different methods on raw images, before automatic white balancing or demosaicing. Because the raw image is captured in extreme low light, it is very dark. For visualisation purposes, we post-process the image by demosaicing using bicubic interpolation and brighten by scaling the image.

For quantitative evaluation, we include PSNR, SSIM~\cite{wang2004image} and Perceptual Index~\cite{wang2018esrgan} (PI). PSNR and SSIM~\cite{wang2004image} are well-known and established measures to assess image quality by comparing the estimated denoised image to a reference image.  For each, a higher number is desired, representing a better match between the estimated denoised image and the reference image.  We augment these quantitative results with those produced by the recently proposed perceptual index~\cite{wang2018esrgan} to better assess the visual quality as perceived by human observers. They evaluate the performance by the non-reference measurement from the PIRM-SR Challenge~\cite{blau20182018}. This index is a linear combination of NIQE~\cite{mittal2013making} and Ma's methods~\cite{ma2017learning} and does not require a reference, i.e. PI $= 0.5*(\mathrm{NIQE}+(10-\mathrm{Ma}))$.  For the perceptual index, a lower score indicates better quality. 

The quantitative results are shown in Table~\ref{tab}. Note that higher PSNR/SSIM~\cite{wang2004image} values are related to the fact that the images are captured under extreme-low light, so the pixel intensities are small in general over a large range (10 bit values). Our long exposure images serving as ground truth also had some defective pixel residual noise, visible in Figure~\ref{fig:synthetic} (left). We observed NODE is effective at removing defective pixels, but PSNR/SSIM will penalise NODE on defective pixels in such cases, therefore we mask out the defective pixels in the ground truth (but not noisy image) using the method described in Section 4.1. These evaluation results are marked as \textit{MASK} in Table~\ref{tab}. Using all metrics, the proposed method outperforms the state-of-the-art. Representative qualitative results are shown in Figure~\ref{fig:qualitative1} and Figure~\ref{fig:qualitative2}.  From the qualitative results, it is apparent that the proposed method can better handle the noise caused by the defective points.  Particularly BM3D struggles with defective pixel correction, as it relies on self-similarity which is less relevant in the presence of defective pixels.  Whilst the deep learning networks do better with the defective pixel noise, there is residual noise contamination that is best removed by NODE.

\textbf{Limitations}
Although NODE demonstrates considerable strength in denoising extreme low light raw images, there are limitations to this research.  First, the proposed method was developed using raw data collected from a single phone model.  In practical setting, this could be feasible approach for producing a targeted denoising method.  However, this paper does not consider generalisation to other phone models, which is left for future work.  Also, the data was collected at a single ISO 12800. However, recent work~\cite{brooks18raw} has shown it is straightforward to include additional input channels to give the denoising network knowledge of the expected noise level. 
\section{Conclusion}
Multi-task noise decomposition proves to be a promising approach for the task of denoising extreme low light raw images.  By letting each subnetwork focus on noise of a particular type, better results can be obtained compared to single-task denoising networks. Future improvements can address adaptation of the method to additional sensors, ISOs, and data types including video.

{\small
\bibliographystyle{ieee}
\bibliography{ICCV2019_Hao_Gregory_v2}
}

\end{document}